#### PHYSICS

# Asymptotic turbulent friction in 2D rough-walled flows

Alexandre Vilquin[1], Julie Jagielka[1], Simeon Djambov[1], Hugo Herouard[1], Patrick Fisher[2], Charles-Henri Bruneau[2], Pinaki Chakraborty[3], Gustavo Gioia[4]*, Hamid Kellay[1]*

The friction $f$ is the property of wall-bounded flows that sets the pumping cost of a pipeline, the draining capacity of a river, and other variables of practical relevance. For highly turbulent rough-walled pipe flows, $f$ depends solely on the roughness length scale $r$, and the $f - r$ relation may be expressed by the Strickler empirical scaling $f \propto r^{1/3}$. Here, we show experimentally that for soap film flows that are the two-dimensional (2D) equivalent of highly turbulent rough-walled pipe flows, $f \propto r$ and the $f - r$ relation is not the same in 2D as in 3D. Our findings are beyond the purview of the standard theory of friction but consistent with a competing theory in which $f$ is linked to the turbulent spectrum via the spectral exponent $\alpha$: In 3D, $\alpha = 5/3$ and the theory yields $f \propto r^{1/3}$; in 2D, $\alpha = 3$ and the theory yields $f \propto r$.

### INTRODUCTION

Wall-bounded flows are unavoidably attended by friction (*1*–*3*). For pipe flows and channel flows, friction may be defined as the dimensionless ratio $f \equiv \tau/(\rho U^2)$, where $\tau$ is the shear stress that develops between the flow and the wall, $\rho$ is the density of the fluid, and $U$ is the mean velocity of the flow. Early empirical research on $f$ started back in the 18th century and appertained to waterways lined with gravel or vegetation, in which the wall is rough and the flow, turbulent (*4*, *5*). By the end of the 19th century, turbulence had been related to large values of the Reynolds number $Re \equiv UW/\nu$ (*6*) (where $W$ is the characteristic width of the flow, a pipe's diameter, for example, and $\nu$ is the kinematic viscosity of the fluid), and much empirical data on rough-walled flows had been encapsulated in the Manning formula (*5*). This prominent empirical formula is customarily used to design canals and pipelines, to estimate the discharge of streams and flood plains, to ascertain the destabilizing effect of water flow in ice sheets (*7*), and to furnish suitable boundary conditions for computational simulations of turbulent rough-walled flows (to cite only a few examples). The Manning formula applies in the highly turbulent asymptotic regime in which $f$ becomes independent of $Re$ as $Re \to \infty$ (*3*, *8*, *9*). For rough-walled pipe flows with a single, finite roughness length scale $r$, the formula simplifies to the Strickler scaling $f \propto (r/W)^{1/3}$ (*8*, *10*), where $W$ is the pipe diameter. At this point, we ask, does the Strickler scaling remain valid for the two-dimensional (2D) equivalent of such rough-walled flows? Or is it that there exists a 2D Strickler scaling distinct from its 3D counterpart?

These questions cannot be answered by invoking the standard theory of friction (*1*–*3*). Nothing in the formulation of that theory pertains to the dimensionality of the flow or could enable separate predictions for 3D and 2D. This limitation of the standard theory becomes compounded if we recall that the dimensionality of a turbulent flow is inextricably tied to the turbulent spectrum (*11*, *12*), which the standard theory ignores.

The turbulent spectrum (*13*) may be said to represent the essential structure of turbulence. It is a function of the wave number $k$, $E(k)$, which can be used to compute the characteristic velocity $u_s$ of a turbulent fluctuation, or "eddy," of size $s$ in the flow (*2*): $u_s \propto (\int_{1/s}^{\infty} E(k)\, dk)^{1/2}$. If $\alpha$ denotes the spectral exponent, then $E(k) \propto U^2 W^{(1-\alpha)} k^{-\alpha}$, and therefore

$$u_s \propto U(s/W)^{(\alpha-1)/2} \quad (1)$$

where $W$ is assumed to be larger than $s$.

A single type of spectrum is possible in 3D flows; known as the "energy cascade," it corresponds to $\alpha = 5/3$ (*14*). Confinement to two spatial dimensions precludes vortex stretching, which is a cardinal trait of 3D turbulence. As a result of this and other disparities between 3D turbulence and 2D turbulence, 2D flows may display a type of spectrum known as the "enstrophy cascade," which corresponds to $\alpha = 3$ (*15*, *16*).

### RESULTS

With the foregoing considerations in mind, we endeavor to measure the $f - r$ relation in 2D. In each experiment, we shall confirm the dimensionality of the flow by determining the spectral exponent.

The 2D equivalent of rough-walled pipe flows may be realized in a soap film (*17*) in which a steady flow becomes established by the action of gravity (Fig. 1A). The thickness of the film, typically 10 µm, is much smaller than both the width of the flow and the roughness length scale (see the Supplementary Materials). Thus, the velocity of the flow lies on the plane of the film, and the flow is 2D.

We pierce the film with a comb, as indicated in Fig. 1A, in order for the flow to turn turbulent as it moves past the teeth of the comb. To visualize the effect of the comb, we cast light on a face of a film and take a snapshot (Fig. 1B) of the changeful interference fringes that form there. The snapshot of Fig. 1B may be interpreted as a contour map of the instantaneous turbulent fluctuations downstream of the comb (*17*).

We compute the spectrum $E(k)$ at numerous points on the film by carrying out measurements using laser Doppler velocimetry (LDV) (see Materials and Methods). From log-log plots of the spectrum, of which a few typical examples may be seen in Fig. 2, it is apparent that the spectral exponent $\alpha$ is always in good accord with the theoretical value for the enstrophy cascade ($\alpha = 3$).

[1]Laboratoire Ondes et Matière d'Aquitaine, UMR 5798 Université Bordeaux et CNRS, 351 cours de la Libération, 33405 Talence, France. [2]Institut de Mathématiques de Bordeaux, UMR 5251 Université Bordeaux et CNRS, 351 cours de la Libération, 33405 Talence, France. [3]Fluid Mechanics Unit, Okinawa Institute of Science and Technology Graduate University, 1919-1 Tancha, Onna-son, Okinawa 904-0495, Japan. [4]Continuum Physics Unit, Okinawa Institute of Science and Technology Graduate University, 1919-1 Tancha, Onna-son, Okinawa 904-0495, Japan.
*Corresponding author. Email: ggioia@oist.jp (G.G.); hamid.kellay@u-bordeaux.fr (H.K.)

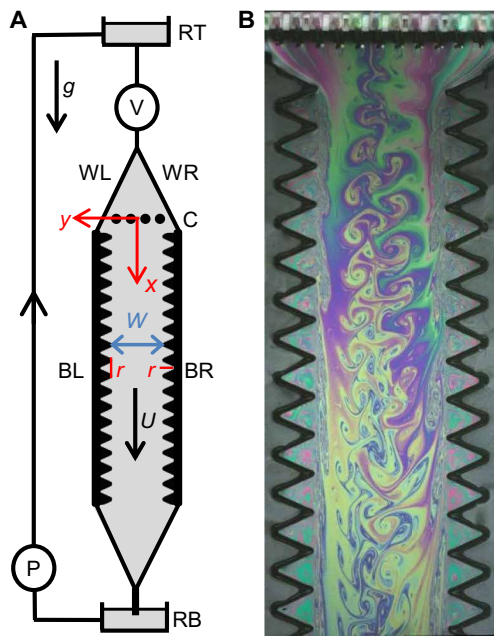

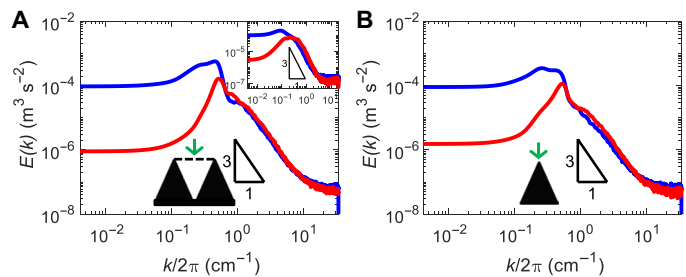

**Fig. 1. Experimental setup used to study soap film flows.** These flows are the 2D equivalent of rough-walled pipe flows with a single roughness length scale $r$. (**A**) Schematic. A soapy solution (2% Fairy Dreft Ultra in water; $v = 0.01$ cm$^2$ s$^{-1}$) drains steadily from reservoir RT through valve V and into the film (drawn in light gray), where it flows driven by gravity "$g$." The film hangs from wires WL and WR (diameter, 0.4 mm) and from thin (thickness = 2 mm), long (length ≈ 1 m), mutually parallel, serrated plastic blades BL and BR (drawn in black), which serve as the rough "walls" of the soap film flow (distance between consecutive serration tips = wavelength of the roughness = amplitude of the roughness = $r$, in the range of 2 to 20 mm). The distance between BL and BR is the width of the flow, $W$ (in the range of 2 to 10 cm). In all cases, $W > r$. After flowing through the film, the soapy solution drains into reservoir RB and returns to RT through pump P. (**B**) Interference fringes evince the presence of 2D turbulent fluctuations downstream of comb C (tooth diameter ≈ 1.5 mm and tooth spacing ≈ 2.5 mm) for a soap film flow with $W = 1.52$ cm and $r = 5$ mm.

**Fig. 2. Typical log-log plots of the turbulent spectrum $E(k)$, from LDV measurements.** $E_{uu}(k_x)$ (blue) and $E_{vv}(k_x)$ (red) are the two realizations of $E(k)$ that can be computed from the LDV measurements, which are carried out at a point equidistant from and ≈1 mm above two consecutive tips of the rough wall (**A**), at a point distant ≈ 1 mm from a tip of the rough wall (**B**), and at a point on the centerline of the flow [inset in (A)]. The flow corresponds to $W = 2$ cm and $r = 1$ cm. All spectra are consistent with turbulence of spectral exponent 3, which, for theoretical reasons, may be expected to decay. Nevertheless, in that part of the flow where we carry out measurements (mostly at a distance of between $8W$ and $20W$ downstream of the comb, where $W$ is the width of the flow), the rate of decay is quite low, and it has a negligible effect on the turbulent friction (see the Supplementary Materials); this remains the case if the comb's tooth spacing is changed from ≈2.5 mm to ≈1 cm.

For any given rough-walled soap film flow of the type sketched in Fig. 1A, we can determine a Reynolds number–friction data point ($Re, f$) by carrying out field measurements using particle image velocimetry (PIV) (see Materials and Methods). These field measurements span a probing section of streamwise length $4r$ (or four times the wavelength of the rough walls of the flow).

In Fig. 3A, we plot the time-averaged streamwise velocity field $u(x, y)$ of a typical flow, and in Fig. 3B, we plot the accompanying time-averaged transversal velocity field $v(x, y)$. By averaging $u(x, y)$ and $v(x, y)$ along the streamwise direction, we obtain the functions $u(y)$ and $v(y)$ of Fig. 3C. The mean velocity of the flow, $U$, may be determined by averaging $u(y)$ along the transverse direction (see Fig. 3C). The Reynolds number follows readily as $UW/v$.

For the same typical flow, we plot the time-averaged total shear stress field $\tau(x, y)$ (Fig. 3D) and the time-averaged turbulent shear stress field $\tau_t(x, y)$ (Fig. 3E) (see the Supplementary Materials). By averaging these stress fields along the streamwise direction, we obtain the functions $\tau(y)$ and $\tau_t(y)$ of Fig. 3F. Function $\tau(y)$ attains peak values $\tau^+$ and $\tau^-$ close to the rough walls, at $y \approx W/2$ and $y \approx -W/2$, respectively (see dashed lines in Fig. 3F). To compute $f \equiv \tau/\rho U^2$, we set $\tau = (|\tau^+| + |\tau^-|)/2$. This completes the determination of a data point ($Re, f$).

Next, we seek to ascertain the relation between $f$ and $Re$ at fixed relative roughness $r/W$. To achieve that goal, we determine, via the procedure exemplified in Fig. 3, the data points ($Re, f$) shown in Fig. 4. Figure 4 indicates that for every given, finite value of $r/W$, $f$ is independent of $Re$, consistent with the high-$Re$ asymptotic regime in which $f$ depends solely on the relative roughness. To study the asymptotic $f - r/W$ relation, we transfer to Fig. 5 every data point of Fig. 4, in the recast form ($r/W, f$).

In addition to the experimental results transferred from Fig. 4, Fig. 5 includes three computational data points (large red circles and green stars in Fig. 5). We obtain these data points by solving the 2D Navier-Stokes equations using direct numerical simulations (see Materials and Methods) in which the rough walls are accounted for by means of a penalty method and in which the flow is made turbulent by piercing the flow with a comb, just as in the experiments (see Fig. 1A). The computational equivalent of Fig. 3 may be found in the Supplementary Materials. Each computational data point comes in two versions. In one version (large red circles in Fig. 5), $f$ is computed as in the soap film experiments, using the peak values of function $\tau(y)$ (see Fig. 3E or its computational equivalent). In the other version (green stars in Fig. 5), $f$ is computed, as is customary in pipe flows, using the formula $f = (\Delta P/\Delta x)(W/2\rho U^2)$, where $\Delta P/\Delta x$ is the streamwise pressure gradient.

From Fig. 5, we conclude that there exists a distinct 2D Strickler scaling, $f \propto r/W$, which cannot be reconciled with its 3D counterpart, $f \propto (r/W)^{1/3}$. This conclusion renders the standard theory incomplete, because that theory can furnish no indication as to what specific difference, if any, is to be expected between the $f - r/W$ relation in 3D and the $f - r/W$ relation in 2D.

On the other hand, the Strickler scalings in 2D and 3D can both be explained by invoking a competing, "spectral theory" of friction in which $f$ is linked to the turbulent spectrum. In the spectral theory, the asymptotic friction of any highly turbulent rough-walled flow with a single roughness length scale $r$ is predicted to be $f \propto u_r/U$ (18, 19), where $u_r$ is the characteristic velocity of a turbulent eddy of size $r$. For a flow of characteristic width $W$ and spectral exponent $\alpha$, $u_r \propto U(r/W)^{(\alpha - 1)/2}$, as may be seen from Eq. 1. It follows that $f \propto (r/W)^{(\alpha-1)/2}$, which is a generalized Strickler scaling (20) for flows of spectral exponent $\alpha$. In 3D, $\alpha = 5/3$ and, thus, $f \propto (r/W)^{1/3}$; in 2D, $\alpha = 3$ and, thus, $f \propto (r/W)$.

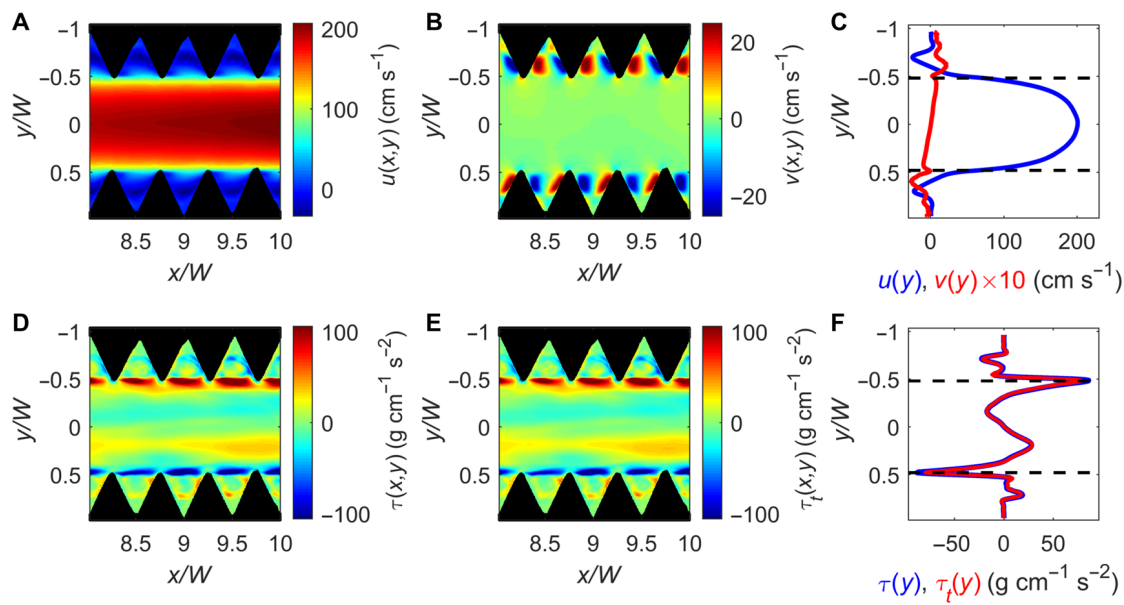

**Fig. 3. Typical velocity fields and shear-stress fields obtained by PIV and used to determine data points (Re, f).** The flow corresponds to $W = 2$ cm and $r = 1$ cm. (**A**) Contour plot of $u(x,y) \equiv \overline{u(x,y,t)}$, where $u(x, y, t)$ is the instantaneous streamwise velocity field at time $t$ and $\overline{(\cdot)}$ denotes time averaging. (**B**) Contour plot of $v(x,y) \equiv \overline{v(x,y,t)}$, where $v(x, y, t)$ is the instantaneous transversal velocity field. (**C**) Plots of $u(y) \equiv (1/4r)\int_0^{4r} u(x,y)\,dx$ (blue) and $v(y) \equiv (1/4r)\int_0^{4r} v(x,y)\,dx$ (red). Function $u(y)$ is used to compute $U = (1/W)\int_{-W/2}^{W/2} u(y)\,dy$ and $Re = UW/\nu$. (**D**) Contour plot of the total shear-stress field $\tau(x, y)$, which is the sum of the viscous shear-stress field $\rho\nu\partial u(x, y)/\partial y$ and the turbulent shear-stress field $\tau_t(x,y) \equiv -\rho\overline{(u'(x,y,t)v'(x,y,t))}$, where $u'(x, y, t) \equiv u(x, y, t) - u(x, y)$ and $v'(x, y, t) \equiv v(x, y, t) - v(x, y)$. (**E**) Contour plot of $\tau_t(x, y)$. (**F**) Plots of $\tau(y) \equiv (1/4r)\int_0^{4r} \tau(x,y)\,dx$ (blue) and $\tau_t(y) \equiv (1/4r)\int_0^{4r} \tau_t(x,y)\,dx$ (red).

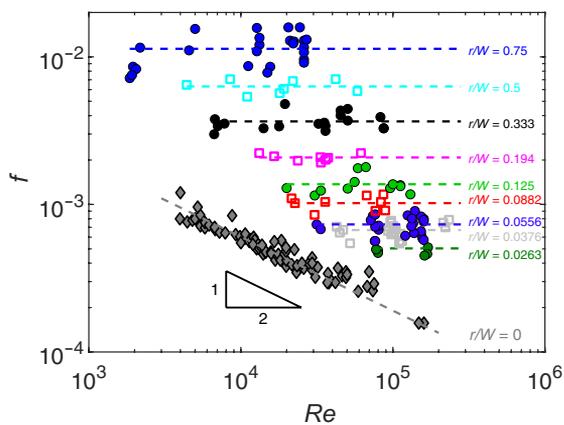

**Fig. 4. Log-log plot of data points (Re, f) for highly turbulent soap film flows of given relative roughness r/W.** For rough-walled flows ($r/W \neq 0$), a horizontal line marks the average friction about which the data points are scattered for each given value of $r/W$. For smooth-walled soap film flows ($r/W = 0$), the empirical data points [taken from (21, 22)] are scattered around the scaling $f \propto Re^{-1/2}$.

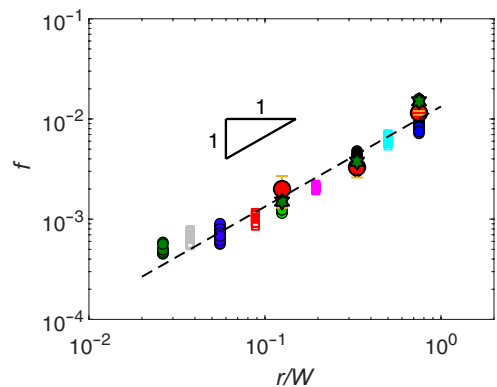

**Fig. 5. Log-log plot of the data points (r/W, f).** The data points are transferred from Fig. 4 (squares and small circles) or computed via the direct numerical simulations described in the main text (large red circles and green stars).

Note that, physically, $f$ is produced by turbulent eddies that transfer momentum between the wall and the flow, and one might expect the larger and faster eddies, of size $\approx W$ and characteristic velocity $\propto U$, to dominate the production of $f$. However, in the spectral theory, it is shown that momentum transfer is dominated by the eddies of size $r$ (18, 19), as may be inferred from the presence of $u_r$ in the expression for $f$.

As has been shown in experiments (21, 22) and simulations (23), the spectral theory can also account for the empirical scalings of highly turbulent smooth-walled flows, namely, $f \propto Re^{-1/4}$, the Blasius scaling, which pertains to 3D, and its 2D counterpart, $f \propto Re^{-1/2}$ (Fig. 4, case $r/W = 0$). The theory predicts the generalized Blasius scaling $f \propto Re^{(1-\alpha)/(1+\alpha)}$ (20, 24), which, for $\alpha = 5/3$ and $\alpha = 3$, yields $f \propto Re^{-1/4}$ and $f \propto Re^{-1/2}$, respectively. [For the 2D case, Falkovich and Vladimirova (25) provide an alternative theoretical explanation of the empirical scaling $f \propto Re^{-1/2}$.]

## DISCUSSION

In view of these results, we submit that the disparity between the turbulent friction in 3D and 2D is a spectral phenomenon. A change in the spectrum brings about a predictable change in the attendant friction, which is the ultimate reason why the Strickler scaling for soap film flows turns out to differ from the Stickler scaling for pipe flows. The same can be said about the Blasius scalings. In rough- and smooth-walled flows alike, friction proclaims the spectrum.

Thus, in the end, the long-standing, standard theory of friction is incomplete because, being predicated on Buckingham's Π theorem and plausible assumptions of similarity (*26*), it entails a noncommital, or indifferent, attitude toward the physical sources of friction. In contrast to the standard theory, the spectral theory singles out the turbulent eddies, and with them the spectrum, as the effective agents of shear-stress production at the interface between the flow and the wall. Spectrum and friction become linked to one another.

It seems likely that the spectral theory will prove useful in meteorology and other disciplines centered on the atmosphere, a sheet-like fluid layer that has long been known to contain large-scale winds with the same spectral exponent we have measured in soap film flows (*12*). Contrariwise, it might be thought that the spectral theory is unlikely to be practically applicable where turbulence is not confined to two spatial dimensions. However, we consider the poorly understood, friction-altering effect of polymer solutes (*27*) (customarily injected into pipelines), suspended sediment (*28*) (naturally occurring in flooding rivers), and other additives [such as air bubbles (*29*)], all of which are known to modify the spectrum of 3D flows in ways that the standard theory is constitutionally incapable of relating to the attendant lessening or increase in turbulent friction. From this example alone, we conjecture that the most promising prospects for the spectral theory will be found in hydraulic engineering, hydrology, geomorphology, and other applied sciences where the flows of interest are invariably 3D.

## MATERIALS AND METHODS
### LDV measurements

For LDV, we seed the soapy solution with polystyrene beads with 1.1 μm in diameter (from Sigma-Aldrich); the bead density is close to that of water. The LDV measurements are carried out at fixed point $(x, y)$ over a time period of about 50 s at a sampling frequency of 20 kHz and yield time series $u(t_i)$ and $v(t_i)$ of the instantaneous velocity of the flow in the streamwise $(x)$ and transversal $(y)$ directions, respectively. To obtain these time series with a uniform spacing, we use linear interpolation; because of the high sampling frequency, the resulting spectrum is insensitive to the order of the interpolation. The attendant fluctuating velocity series may be readily obtained as $u'(t_i) \equiv u(t_i) - \bar{u}$ and $v'(t_i) \equiv v(t_i) - \bar{v}$, where $\bar{u}$ and $\bar{v}$ are the average values of $u(t_i)$ and $v(t_i)$, respectively. To compute the local turbulent spectra, we invoke Taylor's frozen turbulence hypothesis (*13*, *30*) and carry out the space-for-time substitution $t \to x/\bar{u}$, which yields the space series $u(x_i) \equiv u'(t_i)$ and $v(x_i) \equiv v'(t_i)$, where $x_i \equiv \bar{u} t_i$. The spectra $E_{uu}(k_x)$ and $E_{vv}(k_x)$ are the square of the magnitude of the discrete Fourier transform of $u'(x_i)$ and $v'(x_i)$, respectively.

### PIV measurements

To carry out field measurements using digital PIV, we seed the soapy solution with polystyrene beads with 1.1 μm in diameter (from Sigma-Aldrich) or 6 μm in diameter (from Dynoseeds); the bead density is close to that of water. We use the 1.1-μm-diameter particles for the experiments with $W < 5$ cm and the 6-μm-diameter particles for the experiments with $W \geq 5$ cm. This is because large values of $W$ entail a larger field of view and, thus, bigger particles are needed for reliable particle tracking. The plane of the soap film is lit with a laser sheet (wavelength, 532 nm) and filmed with a fast camera (Phantom V641) at a rate of several thousand images per second. Typical spatial resolution is ≈1000 pixels $W^{-1}$. These images (a typical example of which is shown in fig. S1A) allow us to compute, via the MATLAB program *PIVlab* (*31*), the instantaneous velocity fields $u(x, y, t)$ and $v(x, y, t)$, for which we use window sizes of 64 pixel by 64 pixel or 32 pixel by 32 pixel, with two or three passes and 50% overlap. From $u(x, y, t)$ and $v(x, y, t)$, we compute the instantaneous fluctuating velocity field of fig. S1B (for example) as well as mean velocity fields (such as that of fig. S2A) and the mean stresses (viscous and turbulent), by averaging over a time period of ≈1 s, or over 20 transit times of the fluid across the images.

We tested the accuracy of the PIV measurements via direct comparisons with LDV measurements. LDV is a local but accurate method that entails many data points (typically >$10^6$) per measurement. A few such comparisons evince a good agreement between PIV measurements and their LDV counterparts (see fig. S2).

### Direct numerical simulations

The roughness of the walls breaks translational invariance, with the implication that one cannot simply solve the 2D Navier-Stokes equations using spectral methods. This difficulty can be overcome using a penalty method that has been shown to give results in very good agreement with experiments carried out in soap film flows (*32*–*34*). The penalized Navier-Stokes equations with parameter $K$ read

$$\partial_t \vec{v} + (\vec{v} \cdot \vec{\nabla}) \vec{v} - \frac{1}{Re} \Delta \vec{v} + \frac{1}{K} \vec{v} + \vec{\nabla} P = 0$$

$$\vec{\nabla} \vec{v} = 0$$

where $\vec{v} = (u, v)$ is the velocity vector, $P$ is the pressure, and $K$ is a nondimensional permeability coefficient that is nominally "infinite" for the fluid (typically $K = 10^{16}$) and "zero" for the rough walls (typically $K = 10^{-8}$),

A message passing interface parallel code was run using several cores, allowing for very fine grids to model the geometry of the roughness elements or serrations (see fig. S3). After initializing the instantaneous velocity field with a Poiseuille flow at the entrance section, we allow the flow to evolve for a number of transits. The smaller the roughness, the more transits are needed to render the flow turbulent. We pierce the flow with a comb, just as we do in the soap film experiments, in order for the flow to turn turbulent as it moves past the comb's teeth. The value of $K$ is the same for the comb's teeth and for the walls. More information about these simulations can be found in (*32*–*34*).

Figures S3 and S4 show an example of the velocity fields and corresponding stress fields from simulations. These velocity fields are similar to the ones measured in the experiments. The total shear stress calculated is used to compute the turbulent friction by the same method used in experiments (Fig. 5).

The simulations also provide the pressure field $P(x, y, t)$, which can be readily used to determine the cross-sectional pressure as a function of $x$ (see fig. S5). The slope of the linear trendline of that function represents the rate of pressure drop along the streamwise direction and furnishes an independent value of the turbulent friction, as explained in the paper.

## SUPPLEMENTARY MATERIALS

Supplementary material for this article is available at http://advances.sciencemag.org/cgi/content/full/7/5/eabc6234/DC1

**Acknowledgments:** We thank the referees for thoughtful reviews and R. T. Cerbus for helpful discussions. **Funding:** This work was supported by the Institut Universitaire de France, the Conseil Régional Nouvelle Aquitaine, and the Okinawa Institute of Science and Technology Graduate University. **Author contributions:** A.V. and H.K. carried out the experiments, assisted by J.J., S.D., and H.H. C.-H.B. and P.F. carried out the simulations. All authors discussed the results. G.G. wrote the manuscript with input from P.C., A.V., and H.K. All authors discussed the manuscript. **Competing interests:** The authors declare that they have no competing interests. **Data and materials availability:** All data needed to evaluate the conclusions in the paper are present in the paper and/or the Supplementary Materials. Additional data related to this paper may be requested from the authors.

Submitted 4 May 2020
Accepted 10 December 2020
Published 29 January 2021
10.1126/sciadv.abc6234

**Citation:** A. Vilquin, J. Jagielka, S. Djambov, H. Herouard, P. Fisher, C.-H. Bruneau, P. Chakraborty, G. Gioia, H. Kellay, Asymptotic turbulent friction in 2D rough-walled flows. *Sci. Adv.* **7**, eabc6234 (2021).

# Science Advances

## Asymptotic turbulent friction in 2D rough-walled flows

Alexandre Vilquin, Julie Jagielka, Simeon Djambov, Hugo Herouard, Patrick Fisher, Charles-Henri Bruneau, Pinaki Chakraborty, Gustavo Gioia and Hamid Kellay